\documentclass{amsart}
\usepackage{graphicx}
\usepackage{amscd}

\newtheorem{theorem}{Theorem}
\theoremstyle{plain}

\newtheorem{conclusion}{Conclusion}

\newtheorem{corollary}{Corollary}

\newtheorem{definition}{Definition}

\newtheorem{lemma}{Lemma}

\newtheorem{remark}{Remark}

\numberwithin{equation}{section}

\begin{document}

\title[Gaussian Approximation of spacings]{Gaussian Approximations and Related Questions for the Spacings process}
\author{Gane Samb LO}
\address{Research address. L.S.T.A., Universit\'{e} Paris VI. T.45/55, E.3. 4, Place
Jussieu, F-75230, Paris C\'{e}dex 05. France.}


\begin{abstract}
\large
All the available results on the approximation of the k-spacings
process to Gaussian processes have only used one approach, that is the Shorack and Pyke's
one. Here, it is shown that this approach cannot yield a rate better than $%
\left( N/\log \log N\right) ^{-\frac{1}{4}}\left( \log N\right) ^{\frac{1}{2}%
}$. Strong and weak bounds for that rate are specified both where k is fixed
and where $k\rightarrow +\infty $. A Glivenko-Cantelli Theorem is given
while Stute's result for the increments of the empirical process based on
independent and indentically distributed random variables is extended to the spacings process. One of the
Mason-Wellner-Shorack cases is also obtained.

\end{abstract}

\maketitle

\noindent \textbf{Nota Bene.} This paper was part of the PhD thesis, Cheikh Anta Diop University, 1991, not yet published in a peer-reviewed journal by August 2014.

\bigskip 

\section{Introduction}

\Large

\noindent The non-overlapping uniform k-spacings are defined by 
\begin{equation*}
D_{i,n}^{k}=U_{ik,n}-U_{\left( i-1\right) k,n},\text{ \ }1\leq i\leq \left[ 
\frac{n+1}{k}\right] =N,
\end{equation*}

\noindent where $0\equiv U_{0,n}\leq U_{1,n}\leq ...\leq U_{n,n}\leq U_{n+1,n}\equiv 1$
are the order statistics of a sequence $U_{1},...,U_{n}$ of independent
random variables (r.v.'s) uniformly distributed on $\left( 0,1\right) $ and $[x]$ denotes the integer part of $x$. The study of these r.v.'s have received a
great amount of attention in recent years (see \cite{2}, \cite{5}, \cite{10}
and \cite{13}). Particularly the related empirical process 
\begin{equation*}
\beta _{N}\left( x\right) =N^{\frac{1}{2}}\left\{ F_{N}\left( x\right)
-H_{k}\left( x\right) \right\} ,0\leq x\leq +\infty ,
\end{equation*}

\noindent  where 
$$
F_{N}\left( x\right) =\#\left\{ i,1\leq i\leq N,\text{ }%
NkD_{i,n}^{k}\leq x\right\} /N
$$ 

\noindent and 
$$
H_{k}\left( x\right) =\int_{0}^{x}\frac{%
t^{k}e^{-t}}{\left( k-1\right) !} dt, \ \ \ x\geq 0.
$$

\noindent plays a fundamental role in many areas in
statistics (see \cite{5}). All its aspects are have described by various
authors.

\bigskip

\noindent (i) For the convergence of statistics based on spacings, it is helpful to
have a Glivenko-Cantelli Theorem for $F_{N}\left( .\right) $. Such results
for the overlapping case are available in \cite{3}.\\

\bigskip

\noindent (ii) The limiting law of the spacings statistics may follow from suitable
approximations of $\beta _{N}$ to Gaussian processes. It is clear that the
better the rates of those approximations are the less restrictive the
conditions on the underlying random variables ($r.v.$). Such approximations also yield
Kolmogorov-Smirov's tests.\\

\bigskip

\noindent (iii) Finally, the oscillation modulus of $\beta _{N}$ has been studied in 
\cite{7}, where is established the weak behaviour of the oscillation moduli of $\beta
_{N}$ is equivalent to that of the empirical process based on a sequence of 
independent and indentically distributed ($i.i.d$) random variables.\\

\noindent Our aim is to give strong versions of weak characterizations of the
oscillation moduli that we have already given in \cite{7}. As to the
approximation of $\beta _{n}$ to Gaussian processes, we will show that the
rate given in \cite{7} is, in fact, a strong one. Our best achievement is
that this rate is the best attainable for the approach used until now and
we provide the corresponding bounds. With respect to \cite{1} and \cite{2}, we
do not let k fixed. We allow it to go to infinity. Finally we give the
Glivenko-Cantelli Theorem for $F_{N}$ with almost the same condition as in 
\cite{3} for the overlapping case.

\bigskip

\section{The Gaussian approximation.}

\bigskip

\noindent Approximations of $\beta _{N}$ to Gaussian processes are available since 
\cite{12}. The best rates among those already given are due to \cite{1} and
to \cite{2}. Among other results, \cite{2} proved the following theorem and
corollary.

\bigskip

\begin{theorem} \label{t1}. There exists a probability space carrying a sequence $U_{1},U_{2},...$ of
independent r.v.'s uniformly distributed on $\left( 0,1\right) $ and a
sequences of Gaussian processes $\left\{ W_{N}\left( x\right) ,0\leq x\leq
+\infty \right\} ,$ $N=1,2,...$ satisfying 
\begin{equation*}
\forall N>1,\text{ }\mathbb{E}\left( W_{N}\left( x\right) W_{N}\left( y\right)
\right) 
\end{equation*}

\begin{equation}
=\min \left( H_{k}\left( x\right) ,\text{ }H_{k}\left( y\right)
\right) -H_{k}\left( x\right) H_{k}\left( y\right) -k^{-1}xyH_{k}^{\prime
}\left( x\right) H_{k}^{\prime }\left( y\right)   \label{S}
\end{equation}

\noindent such that 

\begin{equation*}
\lim_{N\rightarrow +\infty }\sup \left( \log N\right) ^{-\frac{3}{4}}N^{%
\frac{1}{4}}\sup_{0\leq x\leq +\infty }\left| \beta _{N}\left( x\right)
-W_{N}\left( x\right) \right| <+\infty ,a.s.
\end{equation*}

\noindent whenever k is fixed. Here $H_{k}^{\prime }\left( x\right) =dH_{k}\left(
x\right) /dx.$
\end{theorem}

\bigskip

\begin{remark} \label {r1} From now on, we will say \underline{according to the wording of Theorem \ref{t1}} at the
place of \underline{There exist a probability space ... such that}.
\end{remark}

\bigskip

\begin{definition} A Gaussian process whose covariance function is given by (\ref{S}) will be
called a Shorack process of parameter k or a k-Shorack process.
\end{definition}

\bigskip

\begin{corollary} \label{c1} According to wording of Theorem \ref{t1}, we have 
\begin{equation*}
N^{\frac{1}{4}}\left( \log N\right) ^{-\frac{1}{2}}\left( \log \log N\right)
^{-\frac{1}{4}}\sup_{0\leq x<+\infty }\left| \beta _{N}\left( x\right)
-W_{N}\left( x\right) \right| =0_{p}\left( 1\right) ,\text{ as }N\rightarrow
+\infty .
\end{equation*}

\noindent This means that $a_{N}^{o}=\left( \log N\right) ^{\frac{3}{4}}N^{-\frac{1}{4}%
}$ is a strong rate of convergence while $a_{N}=\left( \log N\right) ^{\frac{%
1}{2}}\left( 2\log \log N\right) ^{\frac{1}{4}}N^{-\frac{1}{4}}$ is a weak
one. In fact \cite{1} has showed
\end{corollary}

\bigskip

\begin{theorem} \label{t2}. There exist another sequence of processes $\beta _{N}^{1},$ $N=1,2,...$
and a sequence of k-Shorack processes $W_{N}^{1},$ $N=1,2,...$ such that,
for k fixed, the two following assertions hold :

\begin{equation*}
\beta _{N}^{1}\text{ } =^{d}\beta _{N},\text{ }\forall N\geq 1 \tag{i}
\end{equation*}

\begin{equation*}
\sup_{0\leq x<+\infty }\left| \beta _{N}^{1}\left( x\right) -W_{N}^{1}\left(
x\right) \right| a.s.=0\left( a_{N}\right)N\rightarrow +\infty, \text{ a.s. }. \tag{ii}
\end{equation*}

\noindent All these results are based on representations of spacings by exponential
r.v.'s. Namely, when $n+1=kN,$%

\begin{equation*}
\left\{ D_{i,n}^{k},1\leq i\leq N\right\}=^{d} \left\{ \frac{\left(
\sum_{j=\left( i-1\right) k}^{j=ik}E_{j}\right) }{S_{n+1}},1\leq i\leq
N\right\}
\end{equation*}

\begin{equation}
=:\left\{ Y_{i}/S_{n+1},1\leq i\leq N\right\}, \label{11}
\end{equation}

\noindent where $E_{1},$ $E_{2},...$ is a sequence of independent exponential rv's
with mean one and whose partial sums are $S_{n},n\geq 1.$ If $\mu
_{N}=\delta _{n}=S_{n+1}/Nk$, it follows that 

\begin{equation*}
\left\{ \beta _{N}\left( x\right) ,0\leq x<+\infty \right\} =^{d}\left\{
N^{\frac{1}{2}}\left( \xi _{N}\mu _{N}\left( x\right) -H_{k}\left( x\right)
\right) +0\left( N^{\frac{1}{2}}\right) \right\}   
\end{equation*}

\begin{equation}
= \left\{ \Lambda_{N}\left( x\right) +R_{N}\left( x\right) ,0\leq
x<+\infty \right\} =:\left\{ \beta _{N}^{\ast }\left( x\right) ,0\leq
x<+\infty \right\}, \label{12}
\end{equation}

\noindent where $\xi _{N}\left( .\right) $ (resp. $\Lambda _{N}\left( .\right) $) is
the empirical distribution function (resp. empirical process) based on $%
Y_{1},...,Y_{N}$. The cited results are derived from simultaneous
approximations of $\Lambda _{N}$ and $R_{N}$.\\

\noindent First, we establish that the best rate attainable through this approach is that
of \cite{1} even when $k\rightarrow +\infty $.
\end{theorem}

\begin{theorem} \label{t3} According to the wording of Theorem \ref{t2}, for any k satisfying 
\begin{equation}
\exists \delta _{0}<0,\text{ }\forall 0<\delta <\delta _{0},kN^{-\delta
}\rightarrow 0\text{ as }N\rightarrow +\infty ,  \tag{L}
\end{equation}

\noindent we have 

\begin{equation*}
\lim_{N\rightarrow +\infty }\sup a_{N}^{-1}\sup_{0\leq x<+\infty }\left|
\beta _{N}^{\ast }\left( x\right) -W_{N}^{\ast }\left( x\right) \right|
a.s.=\left\{ 
\begin{array}{c}
K\left( k\right) =\left( k^{k+\frac{1}{2}}e^{-k}/k!\right) ^{\frac{1}{2}},%
\text{ (k fixed)} \\ 
K_{0}=\left( 2\pi \right) ^{-\frac{1}{4}},\text{ }\left( k\rightarrow
+\infty \right).
\end{array}
\right. 
\end{equation*}

\bigskip

\noindent Our second result is an improvement of Theorem 1 of \cite{2}.
\end{theorem}

\bigskip 

\begin{theorem} \label{t4}. According to the wording of Theorem \ref{t1}, we have for any $k$ such that for
some $\delta _{0},0<\delta _{0}<\frac{1}{4},kN^{-\frac{1}{4}+\delta
_{0}}\rightarrow 0$ as $N\rightarrow +\infty ,$%
\begin{equation*}
\lim_{N\rightarrow +\infty }\sup a_{N}^{-1}\sup_{0\leq x<+\infty }\left|
\beta _{N}\left( x\right) -W_{N}\left( x\right) \right| \leq \left\{ 
\begin{array}{c}
K\left( k\right) ,\text{ (k fixed)} \\ 
K_{0}\text{ \ (k}\rightarrow +\infty \text{)}
\end{array}
a.s.,\right. 
\end{equation*}
\end{theorem}

\bigskip 

\noindent \textbf{Proof of Theorem \ref{t4}}. From (\ref{12}), we have $\beta _{N}$ $=^{d}$ $\beta _{N}^{\ast }$ for all $%
N\geq 1$. Furthermore,
\begin{eqnarray*}
\beta _{N}^{\ast }\left( x\right)  &=&\Lambda _{N}\left( x\right) +N^{\frac{1%
}{2}}\left( H_{k}\left( \mu _{N}x\right) -H_{k}\left( x\right) \right)
-\left\{ \Lambda _{N}\left( \mu _{N}x\right) -\Lambda _{N}\left( x\right)
\right\} +0\left( N^{-\frac{1}{2}}\right)  \\
&=&:\Lambda _{N}\left( x\right) +R_{N1}\left( x\right) +R_{N2}\left(
x\right) +R_{N3}\left( x\right) .
\end{eqnarray*}

\noindent We shall proceed by steps, approximating each of the $R_{Ni}$'s$.$

\bigskip 

\begin{lemma} \label{l1} Let $N_{p}=\left[ \left( 1+\rho \right) ^{p}\right] ,$ $p>0,$ $%
p=1,2,...,\varepsilon >0$ and
\begin{equation*}
C_{N_{p}}=\bigcup_{N=N_{p}}^{N=N_{p+1^{-1}}}\left\{ \sup_{0\leq x<+\infty
}\left| R_{N1}\left( x\right) -N^{\frac{1}{2}}\text{ }xH_{n}^{\prime }\left(
x\right) \left( \mu _{N}-1\right) \right| >\varepsilon a_{N}K\left( k\right)
/4\right\} .
\end{equation*}

\noindent Then if $k/N\rightarrow 0$ as $N\rightarrow +\infty $, $\sum_{p}\mathbb {P}\left(
C_{N_{p}}\right) <+\infty $.
\end{lemma}

\bigskip 

\noindent \textbf{Proof of Lemma \ref{l1}} Apply the mean value theorem twice and get
\begin{equation}
A_{N1}=R_{N1}\left( x\right) -N^{\frac{1}{2}}\left( \mu _{N}-1\right)
xH_{k}^{\prime }\left( x\right) =N^{\frac{1}{2}}\left( \mu _{N}-1\right)
^{2}x^{2}H_{k}^{\prime \prime }\left( x_{N}\right) ,  \label{13}
\end{equation}

\noindent Where $0<\left| x_{N}/x\right| <\max \left( 1,\mu _{N}\right) $. First, it
may be easily seen that
\begin{equation}
\sup_{0\leq x<+\infty }\frac{xH_{k}^{\prime }\left( x\right) }{k^{\frac{1}{2}%
}}=\frac{k^{\frac{1}{2}+k}e^{-k}}{k!}=K\left( k\right) ^{2},  \label{14}
\end{equation}

\begin{equation}
\lim_{k\rightarrow +\infty }\sup_{0\leq x<+\infty }\left| x\text{ }%
H_{k}^{\prime }\left( x\right) /k^{\frac{1}{2}}\right| =K_{0}^{2},
\label{15}
\end{equation}

\noindent and

\begin{equation}
0<M=\sup_{k\geq 1}\sup_{0\leq x<+\infty }\left| x^{2}H_{k}^{\prime \prime
}\left( x\right) /k\right| <+\infty .  \label{16}
\end{equation}

\noindent Recall that for all $\varepsilon >0$, 
\begin{equation}
\sum_{p}\mathbb {P}\left( \max \left( 1,\mu _{N}\right) >1+\varepsilon \right) \leq
\sum_{N}\mathbb {P}\left( \left| \mu _{N}\right| >1+\varepsilon \right) <+\infty, 
\label{17}
\end{equation}

\noindent by the strong law of large numbers (SLLN) and
\begin{equation}
\sum_{p}\mathbb {P}\left( \bigcup_{N=N_{p}}^{N=N_{p+1^{-1}}}\left( \frac{Nk}{2\log
\log nk}\right) \left| \mu _{N}-1\right| >1+\varepsilon \right) <+\infty 
\label{18}
\end{equation}

\noindent by the law of the iterated logarithm (loglog-law). We show in the Appendix
how to adapt the classical SLLN and loglog-law to these cases.\\

\noindent Now by (\ref{13}), (\ref{14}) and (\ref{15}) 

\begin{equation*}
\mathbb {P}\left( C_{N_{p}}\right) \leq \sum_{N=N_{p}}^{N=N_{p+1^{-1}}}\mathbb {P}\left( \max
\left( 1,\mu _{N}\right) ^{2}>1+\varepsilon \right)
\end{equation*}

\begin{equation}
+\mathbb {P}\left(
\bigcup_{N=N_{p}}^{N=N_{p+1^{-1}}}\left| \mu _{N}-1\right| ^{2}\left( \frac{%
Nk}{2\log \log Nk}\right) >ce_{N}\right) ,  \label{19}
\end{equation}

\noindent with $c=\varepsilon K\left( k\right) ^{2}/4M\left( 1+\varepsilon \right)
,e_{N}=\left( \log \log N\right) ^{\frac{1}{4}}N^{\frac{1}{4}}\left( \log
N\right) ^{\frac{1}{2}}\left( 2\log \log Nk\right) ^{-\frac{1}{2}}$. But $%
\log \log Nk=\left( \log N\right) \left( 1+o\left( 1\right) \right) ,$ $%
K\left( k\right) $ is bounded and thus $ce_{N}>\left( 1+\varepsilon \right)
^{2}$ for large N. Thus we can apply (\ref{17}) and (\ref{18}) to (\ref{19})
and this completes the proof.

\begin{lemma} \label{l2} Let $\varepsilon >0$ and 
\begin{equation*}
D_{N_{p}}=\left\{ \bigcup_{N=N_{p}}^{N=N_{p+1^{-1}}}\left( \sup_{0\leq
x<+\infty }\left| R_{N2}\left( x\right) \right| >\left( 1+\varepsilon
/4\right) a_{N}\text{ }K\left( k\right) \right) \right\} ,\text{ }p=1,2,...
\end{equation*}

\noindent Then for any $k=k\left( N\right) $ such that $k/N\rightarrow 0$ as $%
N\rightarrow +\infty ,$ $\sum_{p}p\left( D_{N_{p}}\right) <+\infty .$
\end{lemma}

\bigskip

\noindent \textbf{Proof of Lemma \ref{l2}} The mean value theorem implies 
\begin{equation}
\left| H_{k}\left( \mu _{N}x\right) -H_{k}\left( x\right) \right| \leq
\left| \mu _{N}-1\right| \text{ }K\left( k\right) ^{2}\max \left( 1,\text{ }%
\mu _{N}\right) k^{\frac{1}{2}}.  \label{120}
\end{equation}

\noindent By proceeding similarly to (\ref{19}), we get 
\begin{equation*}
\mathbb {P}\left( D_{N_{p}}\right)  \leq \sum_{N=N_{p}}^{N=N_{p+1^{-1}}}\mathbb {P}\left( \max
\left( 1,\mu _{N}\right) >\left( 1+\varepsilon /4\right) ^{1/3}\right) 
\end{equation*}

\begin{equation*}
+\mathbb {P}\left( \bigcup_{N=N_{p}}^{N=N_{p+1^{-1}}}\left\{ \sup_{\left|
H_{k}\left( x\right) -H_{k}\left( y\right) \right| <c_{N}}\left| \Lambda
_{N}\left( x\right) -\Lambda _{N}\left( y\right) \right| >\left(
1+\varepsilon /4\right) a_{N}K\left( k\right) \right\} \right)  
\end{equation*}

\begin{equation}
=R_{N21}+R_{N22}, \label{121}
\end{equation} 

\noindent with $c_{N}=K\left( k\right) ^{2}k^{\frac{1}{2}}\left| \mu _{N}-1\right|
\left( 1+\varepsilon /4\right) ^{1/3}$. Now, 

\begin{equation}
R_{N22} \leq \mathbb{P}\left( \bigcup_{N=N_{p}}^{N=N_{p+1^{-1}}}\left\{ \left| \mu
_{N}-1\right| >\left( 1+\varepsilon /4\right) ^{1/3}\left( \frac{2\log \log N%
}{Nk}\right) ^{\frac{1}{2}}\right\} \right) \label{122}
\end{equation}

\begin{equation*}
+\mathbb{P}\left( \bigcup_{N=N_{p}}^{N=N_{p+1^{-1}}}\left\{ \sup_{\left|
H_{k}\left( x\right) -H_{k}\left( y\right) \right| \leq b_{N}}\left| \Lambda
_{N}\left( x\right) -\Lambda _{N}\left( y\right) \right| >\left(
1+\varepsilon /4\right) a_{N}K\left( k\right) \right\} \right), 
\end{equation*}

\noindent where $b_{N}=\left( \frac{2\log \log \text{ }N}{N}\right) ^{\frac{1}{2}%
}K\left( k\right) ^{2}\left( 1+\varepsilon /4\right) ^{2/3}$. Let $\gamma
_{N}\left( .\right) $ be the empirical process based on $U_{1},...,U_{N}$
and $P_{N_{p}}$ be the second term of the right member of the inequality (%
\ref{122}). Thus (\ref{11}) implies 

\begin{equation}
P_{N_{p}}\leq \mathbb {P}\left( \bigcup_{N=N_{p}}^{N=N_{p+1^{-1}}}\left\{ \sup_{0\leq
u\leq 1-b_{N}}\frac{\gamma _{N}\left( u\right) -\gamma _{N}\left(
u+b_{N}\right) }{\left( 2b_{N}\log b_{N}^{-1}\right) ^{\frac{1}{2}}}%
>1+\varepsilon _{1}\right\} \right),  \label{123}
\end{equation}

\noindent $1+\varepsilon _{1}<\left(1+\varepsilon \right) ^{2/3}$, where we have used the fact that $\left( 2b_{N}\log b_{N}^{-1}\right) ^{\frac{1}{2}}/a_{N}k\left( k\right) \rightarrow \left( 1+\varepsilon \right)
^{1/3}$ as $k/N\rightarrow 0$, as $N\rightarrow +\infty $. Finally, from
line 14, p.95 and line 23, p.98 in \cite{13}, we get $\sum_{p}P_{N_{p}}<+%
\infty $. This and (\ref{120}), (\ref{121}), (\ref{122}) and (\ref{123})
together imply Lemma \ref{l2}.

\bigskip

\begin{lemma} \label{l3} (Koml\'os, M\'ajor,Tusn\'ady, 1975). There exist a probability space carrying a
sequence $Y_{1},Y_{2},...$ as defined in (\ref{11}) and a sequence of
Brownian bridges 
\begin{equation*}
B_{N}^{1}\left( s\right) ,0\leq s\leq 1,\text{ }N=1,2,...
\end{equation*}

\noindent such that 
\begin{equation*}
\forall N\geq N_{1},\mathbb {P}\left( \sup_{0\leq x<+\infty }\left| \Lambda _{N}\left(
x\right) -B_{N}^{1}\left( H_{k}\left( x\right) \right) \right| >\frac{A\log 
\text{ }N+x}{N^{\frac{1}{2}}}\right) \leq Be^{-\lambda x},
\end{equation*}

\noindent for all sequence $\left( k=k\left( N\right) \right) _{N\geq 1}$ and for all
x, where $N_{1},A,B$ and $\lambda $ are absolute positive constants.
\end{lemma}

\bigskip

\noindent \textbf{Proof of Lemma \ref{l3}} This doesn't need to be proved. It is directly derived from \cite{6} and
Corollary 4.4.4 of \cite{4}.

\bigskip

\noindent \textbf{Proof of Theorem \ref{t3} continued.} On the probability space of Lemma \ref{l3}, Lemmas \ref{l1} and \ref{l2} combined with the fact $%
R_{N3}\leq N^{-\frac{1}{2}}$ imply that 

\begin{equation}
\sum_{p}\mathbb {P}\left( \bigcup_{N=N_{p}}^{N=N_{p+1^{-1}}}\sup_{0\leq x\leq +\infty
}\left| \beta _{N}^{\ast }\left( x\right) -\beta _{N}^{\ast \ast }\left(
x\right) \right| >\left( 1+3\varepsilon /4\right) a_{N}K\left( k\right)
\right) <+\infty ,  \label{124}
\end{equation}

\noindent where $\beta _{N}^{\ast \ast }\left( x\right) =\Lambda _{N}\left( x\right)
-N^{\frac{1}{2}}$ x $H_{k}^{\prime }\left( x\right) \left( \mu _{N}-1\right)
,0\leq x<+\infty $. Hence, the proof will be complete if we approximate $%
\beta _{N}^{\ast \ast }$ in the right way. But by Lemma \ref{l3}, for any $%
\varepsilon >0$, for large N 
\begin{equation}
\mathbb {P}\left( \sup_{0\leq x<+\infty }\left| \Lambda _{N}\left( x\right)
-B_{N}^{1}\left( H_{k}\left( x\right) \right) \right| >A_{1}\left( \log
N\right) ^{2}N^{-\frac{1}{2}}\right) \leq N^{-1-\varepsilon },  \label{125}
\end{equation}

\noindent where $A_{1}$ is some absolute constant. From Lemma 3.1 of \cite{2} 

\begin{equation*}
N^{\frac{1}{2}}\left( \mu _{N}-1\right) =N^{\frac{1}{2}}k\frac{S_{n+1}-Nk}{Nk%
}+k^{-1}\int_{0}^{+\infty }\left\{ \Lambda _{N}\left( x\right)
-B_{N}^{1}\left( H_{k}\left( x\right) \right) \right\} dx
\end{equation*}

\begin{equation}
+k^{-1}\int_{0}^{+\infty }B_{N}^{1}\left( H_{k}\left( x\right) \right) dx. \label{126}
\end{equation}

\noindent Let $t_{N}=N^{\frac{1}{4}-\delta },0\leq \delta \leq \delta _{0}$. On the
one hand, one has for large N.

\begin{equation*}
\mathbb {P}\left( \left| \int_{0}^{t_{N}}\left\{ \Lambda _{N}\left( x\right)
-B_{N}^{1}\left( x\right) \right\} dx\right| >\varepsilon a_{N}/12\right) 
\end{equation*}

\begin{equation*}
\leq \mathbb {P}\left( \sup_{0\leq x<+\infty }\left| \Lambda _{N}\left( x\right)
-B_{N}^{1}\left( H_{k}\left( x\right) \right) \right| >\frac{\varepsilon
\left( 2\log \log N\right) ^{\frac{1}{4}}\left( \log N\right) ^{\frac{1}{2}}%
}{12N^{\frac{1}{4}-\delta }}\right)
\end{equation*}

\begin{equation*}
\leq \mathbb {P}\left( \sup_{0\leq x<+\infty }\left| \Lambda _{N}\left( x\right)
-B_{N}^{1}\left( H_{k}\left( x\right) \right) \right| >A_{1}\log N/N^{\frac{1%
}{2}}\right).
\end{equation*}

\noindent This and (\ref{15}) together imply 

\begin{equation}
\mathbb {P}\left( \sup_{0\leq x<+\infty }\left| xH_{k}^{\prime }\left( x\right)
k^{-1}\int_{0}^{t_{N}}\left\{ \Lambda _{N}\left( t\right) -B_{N}^{1}\left(
H_{k}\left( t\right) \right) \right\} dt\right| >\varepsilon a_{N}K\left(
k\right) /12\right) \leq N^{-1-\varepsilon },  \label{127}
\end{equation}

\noindent for N large enough. On the other hand, as $N\rightarrow +\infty ,$%

\begin{equation}
\mathbb {P}\left( \sup_{0\leq x<+\infty }\left| \int_{t_{N}}^{+\infty }\left\{ \Lambda
_{N}\left( t\right) -B_{N}^{1}\left( H_{k}\left( t\right) \right) \right\}
dt\right| >N^{-\frac{1}{2}}\right) \leq N^{\frac{1}{2}}\exp \left( -N^{\frac{%
1}{4}-\delta }/4\right) .  \label{128}
\end{equation}

\noindent To see that, apply Markov's inequality with 
\begin{equation*}
\mathbb {E}\int_{t_{N}}^{+\infty }\left| \Lambda _{N}\left( x\right) -B_{N}^{1}\left(
H_{k}\left( x\right) \right) \right| dx\leq \int_{t_{N}}^{+\infty
}4k^{-1}e^{-x/2}\frac{x^{\left( k-1\right) /2}}{\left( k-1\right) !}dx\leq
4k^{-1}t_{N}^{k}\exp \left( -t_{N}/2\right) .
\end{equation*}

\noindent Since $k=o\left( N^{\frac{1}{4}-\delta }o\right) $, as $N\rightarrow +\infty 
$, (\ref{128}) follows. Finally for large N, 

\begin{equation*}
\mathbb {P}\left( \sup_{0\leq x<+\infty }\left| xH_{k}^{\prime }\left( x\right) /k^{%
\frac{1}{2}}\frac{S_{Nk}-S_{n+1}}{\left( Nk\right) ^{\frac{1}{2}}}\right|
>\varepsilon a_{N}K\left( k\right) /16\right)
\end{equation*}

\begin{equation*}
\leq \mathbb {P}\left( S_{k}>N^{\frac{1}{%
2}}k^{\frac{1}{2}}\right) =1-H_{k}\left( N^{\frac{1}{2}}k^{\frac{1}{2}%
}\right).
\end{equation*}

\noindent Integrating by parts we have : $k/x\leq \frac{1}{2}\Rightarrow 1-H_{k}\left(
x\right) \leq 2x^{k-1}e^{-x}/\left( k-1\right) !$. Then if $k/N\leq \frac{1}{%
2}$ for large N, we get by Sterling's formula, 
\begin{equation}
1-H_{k}\left( k^{\frac{1}{2}}N^{\frac{1}{2}}\right) \leq const.\text{ }\exp
\left( -k^{\frac{1}{2}}N^{\frac{1}{2}}\left( 1+\left( k/N\right) ^{\frac{1}{2%
}}\log \left( k/N\right) \right) \right) .  \label{129}
\end{equation}

\noindent Thus, 

\begin{equation}
\mathbb {P}\left( \sup_{0\leq x<+\infty } xH_{k}^{\prime }\left( x/k\right)
\left( \left( S_{Nk}-S_{n+1}\right) /N^{\frac{1}{2}}\right) >\varepsilon
a_{N}K\left( k\right) /12\right)
\end{equation}

\begin{equation}
\leq const.\text{ }\exp \left( -\frac{1}{4}%
k^{\frac{1}{2}}N^{\frac{1}{2}}\right),  \label{130}
\end{equation}

\noindent ultimately as $N\rightarrow +\infty $ whenever $k/N\rightarrow 0$ as $%
N\rightarrow +\infty $. Put together (\ref{125}), (\ref{126}), (\ref{127}), (%
\ref{128}) and (\ref{130}) to get 
\begin{equation}
\sum_{N}\mathbb {P}\left( \sup_{0\leq x<+\infty }\left| \beta _{N}^{\ast \ast }\left(
x\right) -W_{N}^{\ast \ast }\left( x\right) \right| >\varepsilon
a_{N}K\left( k\right) /4\right) <+\infty ,  \label{131}
\end{equation}

\noindent where $W_{N}^{\ast \ast }\left( x\right) =B_{N}^{1}\left( H_{k}\left(
x\right) \right) -xk^{-1}H_{k}^{\prime }\left( x\right) \int_{0}^{+\infty }t$
$dB_{N}^{1}\left( H_{k}\left( t\right) \right) ,$ $x\geq 0$. And combine (%
\ref{124}) with (\ref{131}) to have
\begin{equation}
\sum_{p}\mathbb {P}\left( \bigcup_{N=N_{p}}^{N=N_{p+1^{-1}}}\left\{ \sup_{0\leq
x<+\infty }\left| \beta _{N}^{\ast }\left( x\right) -W_{N}^{\ast \ast
}\left( x\right) \right| >\left( 1+\varepsilon \right) a_{N}K\left( k\right)
\right\} <+\infty \right) .  \label{132}
\end{equation}

\noindent This together with Lemma 4.4.4. of \cite{4} completes the proof.

\bigskip 

\begin{proof} of Theorem \ref{t3}. As in the proof of Theorem \ref{t4}, the spacings are always defined on the
probability space of Lemma \ref{l3}. We shall study each of the $R_{Ni}$'s once
again. First we put together (\ref{13}), (\ref{14}), (\ref{15}) and (\ref{16}%
) to get

\begin{equation}
\sup_{0\leq x<+\infty }\left| R_{N1}\left( x\right) -N^{\frac{1}{2}}\left(
\delta _{n}-1\right) xH_{k}^{\prime }\left( x\right) \right| =0\left( N^{-%
\frac{1}{2}}\log \log N\right) ,a.s.,\text{ as }N\rightarrow +\infty .
\label{133}
\end{equation}

\noindent Now Lemma \ref{l2} says nothing else but
\begin{equation}
\lim_{N\rightarrow +\infty }\sup \sup_{0\leq x<+\infty }\left| R_{N2}\left(
x\right) /a_{N}\right| \leq K\left( k\right) \text{ or }K_{0},a.s.,
\label{134}
\end{equation}

\noindent whenever k is fixed or $k\rightarrow +\infty $ while $k/N\rightarrow 0$ as $%
N\rightarrow +\infty $. And the proof will be completed through our
fundamental Lemma which is the following.
\end{proof}

\bigskip 

\begin{lemma} \label{l4} Under the assumptions of Theorem \ref{t3}, we have
\begin{equation*}
\lim_{N\rightarrow +\infty }\sup \sup_{0\leq x<+\infty }\left|
a_{N}^{-1}R_{N2}\left( x\right) \right| \geq K\left( k\right) \text{ or }%
K_{0},a.s.,
\end{equation*}

\noindent according whether k is fixed or $k\rightarrow +\infty $ and satisfies $(L)$.
\end{lemma}

\bigskip 

\begin{proof} of Lemma \ref{l4}. 

\noindent Let $\psi \left( x\right) =\left( \left( k-1\right) !\right)
^{-1}x^{k}e^{-x},$ $x\geq 0$. By the mean value theorem, 
\begin{equation*}
\left| \psi \left( x\right) -\psi \left( k\right) \right| \leq
he^{h}k^{k-1}\left( 1-1/k\right) ^{k-1}\left( \left( k-1\right) !\right)
^{-1},\text{ if }\left| x-k\right| \leq h\leq 1\text{.}
\end{equation*}

\noindent By Sterling's formula we can find a constant $\tau >0$ such that
\begin{equation}
\sup_{\left| x-k\right| \leq h\leq 1}k^{\frac{1}{2}}\left| \psi \left(
x\right) -\psi \left( k\right) \right| \leq \tau hk^{-1},\text{ for all }%
k\geq 1.  \label{135}
\end{equation}

\noindent Now,
\begin{equation}
A_{N}\left( x\right) =H_{k}\left( \delta _{n}x\right) -H_{k}\left( x\right)
=\left( \delta _{n}-1\right) \psi \left( x_{n}\right) \left( x_{n}/x\right)
,0\leq x_{n}/x\leq \max \left( 1,\delta _{n}\right) .  \label{136}
\end{equation}

\noindent If $\left| x-k\right| \leq h\leq 1,\left| x_{n}-k\right| \leq k+\left(
k+h\right) \left| 1-\delta _{n}\right| ,$ and thus by (\ref{135}),

\begin{equation*}
\left| x-k\right| \leq h\leq 1\Rightarrow A_{N}\left( x\right) =\left(
1+o\left( 1\right) \right) k^{\frac{1}{2}}\left( \delta _{n}-1\right)
\end{equation*}

\begin{equation*}
\times \left\{ K\left( k\right) +0\left( \left\{ h+\left( h+k\right) \left|
1-\delta _{n}\right| \right\} /k\right) \right\} ,a.s.
\end{equation*}

\noindent Let $h=h\left( N\right) \rightarrow 0$ as $N\rightarrow +\infty $. Then by
the loglog-law, there exists $\Omega ^{1}\subset \Omega $ and a sequence $%
\left( N_{j\left( \omega \right) }\right) $ extracted from $\left( N\right) $
(let $n_{j}$ and $k_{j}$ be the corresponding subsequences) satisfying

\begin{equation*}
\mathbb {P}\left( \Omega ^{1}\right) =1,\text{ }\forall \omega \in \Omega ^{1},\text{ }%
A_{N_{j}}\left( x\right) =\left( \left( 2\log \log n_{j}\right)
/N_{j}\right) ^{\frac{1}{2}}K\left( k_{j}\right) ^{\frac{1}{2}}\left(
1+o\left( 1\right) \right)
\end{equation*}

\begin{equation}
=:\left( 1+o\left( 1\right) \right) d_{N_{j}}, \label{137}
\end{equation}

\noindent uniformly in $x,$ $k_{j}-h_{j}\leq x\leq k_{j}-h_{j}$, where $h_{j}=h\left(
N_{j}\right) $ as $N\rightarrow +\infty $. Thus we have uniformly un $x$ $ \in
\left[ k_{j}-h_{j},\text{ }k_{j}+h_{j}\right] =I_{k},$%
\begin{equation}
\left| R_{N_{j^{2}}}\left( x\right) \right| d_{=}\left| \gamma
_{N_{j}}\left( H_{k_{j}}\left( x\right) +d_{N_{j}}\left( 1+o\left( 1\right)
\right) -\gamma _{N_{j}}\left( H_{k_{j}}\left( x\right) \right) \right)
\right| =:\left| R_{N_{j^{2}}}^{\ast }\left( x\right) \right| .  \label{138}
\end{equation}

\noindent We now prove that
\begin{equation}
\exists \Omega \subset \Omega ^{1},\text{ }\mathbb {P}\left( \Omega _{0}\right) =1,%
\text{ }\forall \omega \in \Omega _{0},\text{ }\lim_{j\rightarrow +\infty
}\inf \sup_{x\in I_{k_{j}}}\left\{ \left| R_{N_{j^{2}}}^{\ast }\left(
x\right) /b\left( d_{N_{j}}\right) \right| \right\} \geq 1,  \label{139}
\end{equation}

\noindent where $b\left( s\right) =\left( 2s\log \log s^{-1}\right) ^{\frac{1}{2}},$ $%
0<s<1$.\\

\noindent \textbf{Proof of (\ref{139})}.

\noindent Let 
\begin{equation*}
C_{N_{1}}\left( p\right) =\sup_{0\leq v\leq d_{N}/p}\sup_{0\leq s\leq
1-v}\left| \gamma _{N}\left( s\right) -\gamma _{N}\left( s+v\right) \right|
/b\left( d_{N}\right) ,\text{ }p\geq 1.
\end{equation*}

\noindent By Theorem 0.2 of \cite{13},

\begin{equation}
\forall p\geq 1,\text{ }\exists \Omega _{p}\subset \Omega \text{, }P\left(
\Omega _{p}\right) =1,\text{ }\forall \omega \in \Omega _{p},\text{ }%
\lim_{N\rightarrow +\infty }\sup C_{N_{1}}\left( p\right) \left( \omega
\right) <p^{-\frac{1}{2}}.  \label{140}
\end{equation}

\noindent Let 
\begin{equation*}
\Omega =\Omega ^{1} \bigcap \bigcup_{p=1}^{p=+\infty }\Omega _{p}.
\end{equation*}

\noindent Obviously $\mathbb {P}\left( \Omega ^{2}\right) =1$. And for any $\omega \in \Omega ^{2}$, $C_{N_{j^{2}}}\left( \omega \right)=$

\begin{equation*}
\sup_{0\leq x<+\infty }\gamma _{N_{j}}\left( H_{k_{j}}\left( x\right)
+d_{N_{j}}\left( 1+o\left( 1\right) \right)-\gamma _{N_{j}}\left(
H_{k_{j}}\left( x\right) +d_{N_{j}}\right)\right) =o\left( b\left( d_{N_{j}}\right)
\right),    \label{141}
\end{equation*}

\noindent This, together with the following, as $\text{ }j \rightarrow +\infty$,

\begin{equation*}
\forall x \in I_{k_{j}},\text{ }R_{N_{j^{2}}}^{\ast }\left( x\right)
=\gamma _{N_{j}}\left( H_{k_{j}}\left( x\right) +d_{N_{j}}\right) -\gamma
_{N_{j}}\left( H_{k}\left( x\right) \right)
\end{equation*}

\begin{equation}
+\gamma _{N_{j}}\left( H_{k_{j}}\left( x\right) +d_{N_{j}}\left( 1+o\left(
1\right) \right) \right) -\gamma _{N_{j}}\left( H_{k_{j}}\left( x\right)
+d_{N_{j}}\right), \label{142}
\end{equation}

\noindent implies that

\begin{equation*}
\sup_{x\in I_{k_{j}}}R_{N_{j^{2}}}^{\ast }\left( x\right)  \geq \sup_{x\in
I_{k_{j}}}\gamma _{N_{j}}\left( H_{k_{j}}\left( x\right) +d_{N_{j}}\right)
-\gamma _{H_{j}}\left( H_{k_{j}}\left( x\right) \right) +o\left( b\left(
d_{N_{j}}\right) \right)    
\end{equation*}

\begin{equation}
\geq :C_{N_{j^{3}}}\left( h\left( N_{j}\right) \right) +o\left( b\left(
d_{N_{j}}\right) \right). \label{143}
\end{equation}

\noindent Now put $J_{k}=H_{k}\left( I_{k}\right) $ and remark that the lenght of $%
J_{k}$ is $\rho \left( J_{k}\right) =2K\left( k\right) ^{2}nk^{-\frac{1}{2}%
}\left( 1+o\left( 1\right) \right).$\\

\noindent For any $p\geq 1,$ choose $h=h\left( N,\text{ }p\right) =h_{p}$ (with $%
h_{j,p}=h\left( N_{j},p\right) $ such that $2K\left( k\right) ^{2}h_{p}k^{-%
\frac{1}{2}}d_{N}^{-1/4p}=1+o\left( 1\right) $, as $N\rightarrow +\infty $.
Thus, $h\rightarrow 0$ as $N\rightarrow +\infty $ when $(L)$ holds. Also 
$m_{N}=\max \left\{ i,\text{ }i\geq 0,\text{ }H_{k}\left( k-h_{p}\right)
+id_{N}\in J_{k}\right\} \rightarrow +\infty $ as $N\rightarrow +\infty $.
Therefore we may use the lines of the proof of Lemma 2.9 of \cite{13} to
conclude that for any $p\geq 1$,

\begin{equation*}
\mathbb {P}\left( D_{N}\right) =\mathbb {P}\left( \max_{1\leq i\leq m_{N}}\left\{ \gamma
_{N}\left( C_{i+1}^{N}\right) -\gamma _{N}\left( C_{i}^{N}\right) \right\}
/b\left( d_{N}\right) \leq \left( 1-1/p\right) ^{\frac{1}{2}}\right)
\end{equation*}

\begin{equation*}
=0\left( N^{\frac{1}{2}}\exp \left( -m_{N}d_{N}^{1-1/2p}\right) \right) ,
\end{equation*}

\noindent as $N\rightarrow +\infty $, where $C_{i}^{N}=H_{k}\left( k-h_{p}\right)
+id_{N},$ $i=1,...,m_{N}$. But $m_{N}d_{N}=\left( 2K\left( k\right)
^{2}h_{p}k^{-\frac{1}{2}}xd_{N}^{-1/4p}\right)
d_{N}^{1/4p}=d_{N}^{1/4p}\left( 1+o\left( 1\right) \right) $. Hence $\mathbb {P}\left(
D_{N}\right) =0\left( d_{N}^{-1/8p}\right) $ for large N. Thus $%
\sum_{N}\mathbb {P}\left( D_{N}\right) <+\infty $, that is
\begin{equation}
\forall p\geq 1,\text{ }\exists \Omega _{p}^{\prime },\text{ }\mathbb {P}\left( \Omega
_{p}^{\prime }\right) =1,\text{ }\forall \omega \in \Omega _{p}^{\prime },%
\text{ }\lim_{N\rightarrow +\infty }\inf C_{N3}\left( h_{p}\right) /b\left(
d_{N}\right) \geq \left( 1-1/p\right) ^{\frac{1}{2}}.
\end{equation}

\noindent Letting 
\begin{equation*}
\Omega _{0}^{\prime }=\Omega ^{2}\bigcup \bigcup_{p=1}^{p=+\infty },
\end{equation*}

\noindent we get $\mathbb {P}\left( \Omega _{0}^{\prime }\right) =1$ and for all $\omega \in
\Omega _{0}^{\prime },$%
\begin{equation}
\lim_{j\rightarrow +\infty }\inf \sup_{x\in I_{j_{k}}}\left|
R_{N_{j^{2}}}^{\ast }\left( x\right) \right| /b\left( d_{N}\right) \geq 1.
\label{145}
\end{equation}

\noindent We have used in (\ref{138}) that representation for commodity reasons as it
has appeared in the proof. The same may be done, step by step, following
Stute's results (see \cite{13}) to get the version of (\ref{145}) for $%
R_{N_{j^{2}}\text{ }}$itself. This remark completes the proof of (\ref{139}).
\end{proof}

\bigskip 

\begin{proof} of Lemma \ref{l4} (Continued). Remark that

\begin{equation}
\lim_{N\rightarrow +\infty }\sup \sup_{0\leq x<+\infty }\left| R_{N2}\left(
x\right) \right| /b\left( d_{N}\right)  \geq \lim_{j\rightarrow +\infty
}\sup \sup_{0\leq x<+\infty }\left\{ \left| R_{N_{j^{2}}}\left( x\right)
\right| /b\left( d_{N_{j}}\right) \right\}
\end{equation}

\begin{equation*}
\geq \lim_{j\rightarrow +\infty }\inf \sup_{0\leq x<+\infty
}R_{N_{j^{2}}}\left( x\right) /b\left( d_{N_{j}}\right) \geq
\lim_{j\rightarrow +\infty }\inf \sup_{x\in I_{k}}R_{N_{j^{2}}}\left(
x\right) /b\left( d_{N_{j}}\right) .
\end{equation*}

\noindent This combined with (\ref{139}) and with the fact that $b\left( d_{N}\right)
=K\left( k\right) a_{N}\left( 1+o\left( 1\right) \right) $ as $N\rightarrow
+\infty $ proves the Lemma \ref{l4}.
\end{proof}

\bigskip 

\begin{conclusion}
It is clear by Theorem \ref{t3}. that the approach used until now cannot yield a
rate better than $a_{N}$. The problem is now : what new approach would be
used to reach, if possible, the very best rate, that of \cite{6} which is $%
N^{-\frac{1}{2}}\log N$.
\end{conclusion}

\bigskip 

\section{The Glivenko-Cantelli Theorem} 

\noindent For the overlapping case, \cite{3} obtained a Glivenko-Cantelli theorem when
the step satisfies $kN^{-1+a}\rightarrow 0$ as $N\rightarrow +\infty $ for
some $0<a<1$. As to the overlapping case only fixed steps have been handled
in \cite{2}. We give the general result in

\bigskip 

\begin{theorem} \label{t5}. Let $k\geq 1$ be fixed or $k\rightarrow +\infty $ while $k/N\rightarrow 0$
as $N\rightarrow +\infty $. Then
\begin{equation*}
\lim_{N\rightarrow +\infty }\sup_{0\leq x<+\infty }\left| F_{N}\left(
x\right) -H_{k}\left( x\right) \right| =0,a.s.
\end{equation*}

\noindent on the probability space where the spacings are defined.
\end{theorem}

\bigskip 

\begin{proof} of Theorem \ref{t5}. We have

\begin{equation*}
\forall N\geq 1,\left\{ F_{N}\left( x\right) -H_{k}\left( x\right) ,0\leq
x<+\infty \right\}
\end{equation*}

\begin{equation}
=^{d}\left\{ \xi _{N}\left( x\right) -H_{k}\left(
x\right) +R_{N4}\left( x\right) +N^{-\frac{1}{2}}R_{N2}\left( x\right)
+0\left( N^{-\frac{1}{2}}\right) ,0\leq x<+\infty \right\} .  \label{21}
\end{equation}

\noindent First, it follows from Lemma \ref{l2} that for all $\varepsilon >0$,
\begin{equation*}
\sum_{p}\mathbb {P}\left( \bigcup_{N=N_{p}}^{N=N_{p+1^{-1}}}N^{-\frac{1}{2}%
}\sup_{0\leq x<+\infty }\left| R_{N2}\left( x\right) \right| >\varepsilon
/4\right) <+\infty .
\end{equation*}

\noindent Next, 
\begin{equation*}
\mathbb {P}\left( \sup_{0\leq x<+\infty }\left| R_{N4}\left( x\right) \right|
>\varepsilon /4\right) \leq \mathbb {P}\left( \left| 1-\mu _{N}\right| k^{\frac{1}{2}%
}K\left( k\right) ^{2}>\varepsilon /4\right) \text{. }
\end{equation*}

\noindent And direct calculations imply that for all $\lambda >1$, we have
\begin{equation*}
\mathbb {P}\left( \left| 1-\mu _{N}\right| k^{\frac{1}{2}}K\left( k\right)
^{2}>\varepsilon /4\right) \leq \mathbb {P}\left( \left| 1-\mu _{N}\right| \left( 
\frac{Nk}{2\log \log Nk}\right) ^{\frac{1}{2}}>\lambda \right) 
\end{equation*}

\noindent for large N. Thus by (\ref{18})
\begin{equation*}
\sum_{p}\mathbb {P}\left( \bigcup_{N=N_{p}}^{N=N_{p+1^{-1}}}\sup_{0\leq x<+\infty
}\left| R_{N4}\left( x\right) \right| >\varepsilon /4\right) <+\infty 
\end{equation*}

\noindent whenever $k/N\rightarrow 0$ as $N\rightarrow +\infty $. Finally,

\begin{equation*}
\mathbb {P}\left( \sup_{0\leq x<+\infty }\left| \xi _{N}\left( x\right) -H_{k}\left(
x\right) \right| >N^{-\frac{1}{4}}\right)  =\mathbb {P}\left( \sup_{0\leq s<1}N^{-%
\frac{1}{2}}\left| \gamma _{N}\left( s\right) \right| >N^{-\frac{1}{4}%
}\right)  
\end{equation*}

\begin{equation*}
\leq 2N\text{ }\max_{0\leq i\leq N}\text{ }\mathbb {P}\left( U_{i,N}-\frac{i}{N}>N^{-%
\frac{1}{4}}-N^{-1}\right) =J_{N},
\end{equation*}

\noindent by the fact that $\gamma _{N}\left( .\right) $ has stationary increments.
Using now a representation of $\gamma _{N}$ by a Poisson process and an
approximation of a Poisson distribution by a Gaussian one 
(see Lemmas 2.7 and 2.9 in \cite{13}) to get for large N that
\begin{equation*}
J_{N}\leq const.\text{ }N^{3/2}\mathbb {P}\left( N\left( 0,1\right) >N^{-\frac{1}{4}%
}const.\right) \leq const.\text{ }N^{5/4}\exp \left( -N^{1/8}\right) .
\end{equation*}

\noindent Thus $\sum_{N}J_{N}<+\infty $. And the proof of Theorem \ref{t5} is now complete.
\end{proof}

\bigskip 

\section{The oscillation moduli} 
\noindent The oscillation modulus of a function $R\left( s\right) $, $0\leq s<1$, is
defined by 
\begin{equation*}
\kappa \left( d,R\right) =\sup_{0\leq h\leq d}\sup_{0\leq s<1-h}\left|
R\left( s+h\right) -R\left( s\right) \right| ,0<d<1\text{.}
\end{equation*}

\noindent That of the empirical process pertaining to $iid$ $rv$'s has been studied
for several choices of $d$ in \cite{9} and \cite{13}. It is remarkable that
the weak versions of all those results are inherited by the reduced spacings
process $\alpha _{N}\left( s\right) =\beta _{N}\left( H_{k}^{-1}\left(
s\right) \right) ,0\leq s<1$, (see \cite{7}). For the strong case, we obtain
these two results.

\bigskip 

\begin{theorem} \label{t6} 

\noindent I. The Stute's case.\\

\noindent If $\left( d_{N}\right) _{N\geq 1}$ is a sequence of non-increasing positive
reals such that
\begin{equation}
Nd_{N}\rightarrow +\infty ,  \tag{S1}
\end{equation}

\begin{equation}
\left( \log d_{N}^{-1}\right) /\left( Nd_{N}\right) \rightarrow 0,
\tag{S2}
\end{equation}

\begin{equation}
\left( \log d_{n}^{-1}\right) /\log \log N\rightarrow +\infty ,  \tag{S3}
\end{equation}

\begin{equation}
\left( 2d_{N}\log d_{N}^{-1}\right) ^{\frac{1}{2}}/a_{N}=:q_{N}/a_{N}%
\rightarrow +\infty ,\text{ as }N\rightarrow +\infty ,  \tag{S4}
\end{equation}

\noindent then for $k\geq 1$ fixed or $k=k\left( N\right) \rightarrow +\infty $ as $%
N\rightarrow +\infty $ and satisfying
\begin{equation}
\exists N_{o},\delta >2,\text{ }\forall N\geq N_{o},0<d_{N}<k^{k\left(
\delta -2\right) }\exp \left( -\frac{1}{2}k^{\delta }\right) .
\end{equation}

\noindent we have $\lim_{N\rightarrow +\infty }\sup \kappa \left( d_{N},\alpha
_{N}\right) /q_{N}=1$ $\ a.s.$\\

\noindent II. A Mason-Wellner-Shorack case.\\

\noindent Let $a_{N}=\alpha \left( \log N\right) ^{-c},\alpha >0,c>0$. Then under the
same assumptions on k used in Part I, we have $\lim_{N\rightarrow +\infty
}\sup \kappa \left( d_{N},\alpha _{N}\right) /q_{N}\leq \left( 1+c\right) ^{%
\frac{1}{2}},a.s.$
\end{theorem}

\bigskip 

\noindent \textbf{Proof  of Part I of Theorem \ref{t6}}. We have by Lemmas \ref{l1} and \ref{l2},

\begin{equation*}
\forall N\geq 1,\left\{ \alpha _{N}\left( s\right) ,0\leq s<1\right\}
d_{=}\left\{ \Lambda _{N}\left( H_{k}^{-1}\left( s\right) \right)
+R_{N5}\left( s\right) +R_{N6}\left( s\right) ,0\leq s<1\right\} 
\end{equation*}

\begin{equation}
=:\left\{ 
\bar{\alpha}_{N}\left( s\right) ,0\leq s<1\right\} ,  \label{30}
\end{equation}

\noindent with 
\begin{equation*}
R_{N5}\left( s\right) =N^{\frac{1}{2}}\left( \mu _{N}-1\right)
H_{k}^{-1}\left( s\right) H_{k}^{\prime }\left( H_{k}^{-1}\left( s\right)
\right) =:N^{\frac{1}{2}}\left( \mu _{N}-1\right) \phi \left( s\right)
,0\leq s<1,
\end{equation*}

\noindent and

\begin{equation}
\sum_{p}\mathbb {P}\left( \bigcup_{N=N_{p}}^{N=N_{p+1^{-1}}}\sup_{0\leq s<1}\left|
R_{N6}\left( s\right) \right| >\left( 1+\varepsilon \right) a_{N}K\left(
k\right) \right) <+\infty,  \label{31}
\end{equation}

\noindent by (\ref{31}) and $(S4)$, we have

\begin{equation}
\sum_{p}\mathbb {P}\left( \bigcup_{N=N_{p}}^{N=N_{p+1^{-1}}}\kappa \left(
d_{N},R_{N6}\right) >\varepsilon q_{N}/3\right) <+\infty .  \label{32}
\end{equation}

\noindent By Lemma A4 in \cite{7}, $\kappa \left( d_{N},\phi \right) =\left( 1+o\left(
1\right) \right) q_{N}^{2}$ as $N\rightarrow +\infty $ for all k satisfying $(S5)$. Thus, by the loglog-law,

\begin{equation*}
\sum_{p}\mathbb {P}\left( \bigcup_{N=N_{p}}^{N=N_{p+1^{-1}}}\kappa \left(
d_{N},R_{N5}\right) >\varepsilon q_{N}/3\right) <+\infty, 
\end{equation*}

\noindent whenever

\begin{equation}
\lim_{N\rightarrow +\infty }k^{-1}d_{N}\log \log \left( 1/d_{N}\right) \log
\log \text{ }Nk=0  \label{33}
\end{equation}

\noindent is satisfied. This obviously follows from $(S1)$, $(S2)$, $(S3)$, $(S4)$ and $(S5)$. By the results of \cite{13} as recalled in (\ref{123}), for $%
\varepsilon >0$,

\begin{equation}
\sum_{p}\mathbb {P}\left( \bigcup_{N=N_{p}}^{N=N_{p+1^{-1}}}\kappa \left(
d_{N},\Lambda _{N}\left( H_{k}^{-1}\right) \right) >\left( 1+\varepsilon
/3\right) q_{N}\right) <+\infty,   \label{34}
\end{equation}

\noindent when $(S1)$, $(S2)$ and $(S3)$ hold. Since $\varepsilon $ is arbitrary
and since $(S1)$ and $(S3)$ imply (\ref{33}), we get
\begin{equation}
\lim_{N\rightarrow +\infty }\sup q_{N}^{-1}\kappa \left( d_{N},\alpha
_{N}\right) \leq 1,a.s.  \label{35}
\end{equation}

\noindent To get the other inequality, define for $0<c_{1}<c_{2}<+\infty ,0<d<1$, for
any function $R\left( s\right) ,0\leq s<1,$%

\begin{equation}
\kappa ^{\prime }\left( d,\text{ }R\right) =\sup_{c_{1}d<u-t<c_{2}d}\left|
R\left( u\right) -R\left( t\right) \right| /\sqrt{u-t},0\leq u,\text{ }t\leq1. \label{36}
\end{equation}

\noindent Let $R_{N}\left( .\right) =R_{N5}\left( .\right) +R_{N6}\left( .\right) $
and $r_{N}\left( .\right) =\Lambda _{N}\left( H_{k}^{-1}\left( .\right)
\right).$ Now remark that for all $\varepsilon _{1}>0,$ there exists $%
\varepsilon _{2}>0$ such that for 

$$a=\left( \left( 1-\varepsilon _{1}\right)
\log d_{N}^{-1}\right) ^{\frac{1}{2}}
$$

\noindent and 

$$
b=\left( \varepsilon _{2}\log
d_{N}^{-1}\right) ^{\frac{1}{2}},
$$

$$
a+b=\left( \left( 1-\varepsilon
_{1}+\varepsilon _{2}+2\left( \varepsilon _{2}\left( 1-\varepsilon
_{1}\right) \right) ^{\frac{1}{2}}\right) \log d_{N}^{-1}\right) ^{\frac{1}{2%
}}=\left( \left( 1-\varepsilon _{3}\right) \log d_{N}^{-1}\right) ^{\frac{1}{%
2}}
$$

\noindent with $\varepsilon _{3}>0$, $\varepsilon _{3},\varepsilon_{2}\rightarrow 0$ as $\varepsilon _{1}\rightarrow 0$. Thus,

\begin{equation*}
\mathbb {P}\left( \kappa ^{\prime }\left( d_{N},\alpha _{N}\right) \leq a\right) 
\leq \mathbb {P}\left( \left\{ \kappa ^{\prime }\left(d_{N},\bar{\alpha}_{N}\right)\leq a\right\} \bigcup \left\{ \kappa ^{\prime }\left( d_{N},R_{N}\right)>b\right\} \right)
\end{equation*}

\begin{equation*}
+\mathbb {P}\left( \left\{ \kappa ^{\prime }\left( d_{N},\bar{\alpha}_{n}\right)
\leq a\right\} \prod \left\{ \kappa ^{\prime }\left( d_{N},R_{N}\right) \leq
b\right\} \right) 
\end{equation*}

\begin{equation}
\leq \mathbb {P}\left( \kappa ^{\prime }\left( d_{N},R_{N}\right) >b\right) +\mathbb {P}\left(
\kappa ^{\prime }\left( d_{N},r_{N}\right) \leq a+b\right), \label{37}
\end{equation}

\noindent By (\ref{31}) and (\ref{32})

\begin{equation}
\sum_{p}\mathbb {P}\left( \bigcup_{N=N_{p}}^{N=N_{p+1^{-1}}}\kappa ^{\prime }\left(
d_{N},R_{N}\right) >b\right) <+\infty ,  \label{38}
\end{equation}

\noindent for all $\varepsilon _{2}>0$. Thus by (\ref{30}), (\ref{36}), (\ref{37}) and
(\ref{38}) and Lemma 2.9 of \cite{13} and some straightforward
considerations, we get $\lim_{N\rightarrow +\infty }\inf \kappa ^{\prime
}\left( d_{N},\alpha _{N}\right) \geq 1,a.s.,$ under $(S1)$, $(S2)$, $(S3)$ and $(S4)$. Letting $c_{1}=c_{2}=1$,

\begin{equation}
\liminf_{N \rightarrow +\infty} \kappa \left( d_{N},\alpha _{N}\right) \geq
\lim_{N\rightarrow +\infty }\inf \kappa ^{\prime }\left( d_{N},\alpha
_{N}\right) \geq 1,a.s. \label{39}
\end{equation}

\noindent (\ref{35}) and (\ref{39}) together complete the proof of Part I of Theorem \ref{t6}.

\bigskip

\noindent \textbf{Proof of Part II of Theorem \ref{t6}}.

\bigskip 

\noindent Here $(S3)$ and $(S4)$ are satisfied. It suffices thus to write
again the proof of the part one where one should use the probability
inequality (\ref{13}) of \cite{9}. It must be noticied that Part III of Theorem
1 in \cite{9} holds for the general case where $a_{N}=\alpha \left( \log
_{N}\right) ^{-c},0<\alpha ,0<c$.

\newpage 

\textbf{APPENDIX. PROOFS OF STATEMENTS (\ref{17}) AND (\ref{18})}

\bigskip
\noindent a) \textbf{Proof of Statement (\ref{17})}.\\

\noindent Tchebychev's inequality yields $\alpha >1$ and $\beta >1$ such that $\mathbb {P}\left(
S_{n}2/n^{2}>1+\varepsilon \right) \leq A_{2}n^{-\alpha }$ and $\mathbb {P}\left(
\left| S_{n}-S_{m\left( n\right) }\right| >n\varepsilon /2\right) \leq
A_{3}n^{-\beta }$ as $n\rightarrow +\infty ,$ where $m\left( n\right) =\max
\left\{ j^{2},j^{2}\leq n,\text{ }j=1,2,...\right\}$. Thus

\begin{equation*}
\mathbb {P}\left( \left| \mu _{N}-S_{n+1}/\left( n+1\right) \right| >\varepsilon
/2\right) +\mathbb {P}\left( S_{n+1}\geq 1+\varepsilon /2\right)  \\
\end{equation*}

\begin{equation}
\leq \mathbb {P}\left( \left| \mu _{N}-S_{n+1}/\left( n+1\right) \right|
>\varepsilon /2\right) +\left( A_{2}+o\left( 1\right) \right) k^{-\alpha
}N^{-\alpha }+\left( A_{3}+o\left( 1\right) \right) k^{-\beta }N^{-\beta }, \label{a1}
\end{equation}

\noindent since $\left( n+1\right) \sim Nk$ as $N\rightarrow +\infty $. Furthermore,
by Tchebychev's inequality,

\begin{equation*}
\mathbb {P}\left( S_{Nk}/Nk-\left( Nk\right) ^{2}>\varepsilon /8\right) \leq
64N^{-3}k^{-3}/\varepsilon ^{2}
\end{equation*}

\begin{equation*}\mathbb {P}\left( S_{k}/\left( Nk\right)
^{2}-\left( Nk\right) ^{2}>\varepsilon /8\right) \leq
64N^{-4}k^{-3}/\varepsilon ^{2}
\end{equation*}

\noindent and 

\begin{equation*}
\mathbb {P}\left( \left| \mu _{N}-S_{n+1}/\left( n+1\right) \right| \geq \varepsilon
/2\right) \leq \mathbb {P}\left( S_{Nk}/Nk>\varepsilon /4\right) +\mathbb {P}\left( S_{k}/\left(
Nk\right) ^{2}>\varepsilon /4\right) .
\end{equation*}

\noindent Hence since $Nk\rightarrow +\infty ,$ $N^{2}k\rightarrow +\infty $ as $%
N\rightarrow +\infty ,$%

\begin{equation}
\sum_{N}\mathbb {P}\left( \left| \mu _{N}-S_{n+1}/n+1\right| >\varepsilon /2\right)
<+\infty .  \label{a2}
\end{equation}

\noindent Thus (\ref{a1}) and (\ref{a2}) together imply (\ref{17}).\\

\bigskip

\noindent \textbf{Proof of (\ref{18})}.\\

\noindent We have
\begin{equation*}
\frac{S_{n+1}-Nk}{\left( 2Nk\log \log Nk\right) ^{\frac{1}{2}}}=\frac{%
S_{n+1}-S_{Nk}}{\left( 2Nk\log \log Nk\right) ^{\frac{1}{2}}}+\frac{%
S_{n+1}-Nk}{\left( 2Nk\log \log Nk\right) ^{\frac{1}{2}}}=:S_{N}^{\prime
}+S_{N}^{\prime \prime }\text{.}
\end{equation*}

\noindent First, since $0\leq \left( n+1\right) -Nk\leq k,$%

\begin{equation*}
\mathbb {P}\left( S_{N}^{\prime }>\varepsilon /2\right) \leq \mathbb {P}\left( S_{k}>\varepsilon
\left( 2Nk\log \log Nk\right) ^{\frac{1}{2}}/2\right)
\end{equation*}

\begin{equation*}
 \leq 1-H_{k}\left( k^{%
\frac{1}{2}}N^{\frac{1}{2}}\right) \leq const.\exp \left( -\frac{1}{4}k^{%
\frac{1}{2}}N^{\frac{1}{2}}\right) 
\end{equation*}

\noindent as $k/N\rightarrow 0$, $N\rightarrow +\infty $ (see Statement (\ref{129})). Thus
\begin{equation}
\sum_{N}\mathbb {P}\left( S_{N}^{\prime }>\varepsilon /2\right) <+\infty .  \label{a3}
\end{equation}

\noindent Now, let 
\begin{equation*}
p=p\left( N\right) =\inf \left\{ j,N>N_{j}\right\} 
\end{equation*}

\noindent and 
\begin{equation*}
q\left( N\right) =\inf \left\{ j,k\left( N\right) >N_{j}=\left[ \left(
1+\rho \right) ^{j}\right] ,\text{ }j=1,2,...\right\} 
\end{equation*}

\noindent Then $N_{p-1}\leq N\leq N_{p},N_{p-1}N_{q-1}\leq NK\leq N_{p}N_{q},\log \log
N_{p}N_{q}=\left( \log \log N_{p}\right) \left( 1+o\left( 1\right) \right) ,$
as $N/k\rightarrow +\infty ,N\rightarrow +\infty ,N_{p+1}/N_{p}\rightarrow
1+\rho $, as $N\rightarrow +\infty $. Thus (see \cite{8}, p.259-262).

\begin{equation*}
\mathbb {P} \left(\bigcup_{N=N_{p}}^{N=N_{p+1^{-1}}}\left\{ S_{N}^{\prime \prime }\geq
1+\varepsilon /2\right\} \right) \leq A_{4}
\mathbb {P}\left(S_{N_{p}N_{q}}>1+\delta\left( \varepsilon ,\rho \right)\left( 2N_{p}loglogN_{p} \right) ^{ \frac{1}{2} } \right)
\end{equation*}

\begin{equation*}
\leq A_{5}p^{-\left( 1+\delta \left( \varepsilon ,\rho \right)
\right) }
\end{equation*}

\noindent as $p\rightarrow +\infty $, for $\rho $ small enough, $\delta \left(
\varepsilon ,\rho \right) >0$. The same holds for $-S_{N}^{\prime \prime }$.
\noindent Thus

\begin{equation}
\sum_{p}\mathbb {P}\left( \bigcup_{N=N_{p}}^{N=N_{p+1^{-1}}}\left( \left|
S_{N}^{\prime \prime }\right| >1+\varepsilon /2\right) \right) <+\infty. \label{a4}
\end{equation}

\noindent Finally (\ref{a3}) and (\ref{a4}) together imply (\ref{18}).

\end{document}